\documentclass[11pt,dvips]{article}
\usepackage{epsfig,times} 
\usepackage{picinpar}
\usepackage{wrapfig}
\usepackage{floatflt}
\makeatletter
\renewcommand{\section}{\@startsection{section}{1}{0in}
	{0.4\baselineskip}{0.1\baselineskip}{\Large\bf}}
\renewcommand{\subsection}{\@startsection{subsection}{2}{0in}
	{0.25\baselineskip}{-\baselineskip}{\large\bf}}
\renewcommand{\subsubsection}{\@startsection{subsubsection}{3}{0in}
	{0.1\baselineskip}{-\baselineskip}{\normalsize\bf}}
\makeatother
\begin{document}

%
\makeatletter\newcommand{\ps@icrc}{
\renewcommand{\@oddhead}{\slshape{OG 4.3.05}\hfil}}
\makeatother\thispagestyle{icrc}
%
%

\begin{center}
%
{\LARGE \bf An Optical Reflector for the CANGAROO-II Telescope}
\end{center}

\begin{center}
%
%
{\bf Akiko Kawachi\footnote[1]{
Institute for Cosmic Ray Research, University of Tokyo,
     Tanashi, Tokyo 188-8502, Japan, 
$^{2}$Department of Physics, Tokyo Institute of Technology, 
        Meguro, Tokyo 152-8551, Japan, 
$^{3}$Department of Physics and Mathematical Physics, University of 
   Adelaide, South Australia 5005, Australia, 
$^{4}$Institute of Space and Astronautical Science,
   Sagamihara, Kanagawa 229-8510, Japan, 
$^{5}$Department of Physics, Yamagata University, 
Yamagata 990-8560, Japan, 
$^{6}$Faculty of Management Information, Yamanashi Gakuin University,  Kofu, 
Yamanashi 400-8575, Japan,  
$^{7}$Department of Physics, Tokai University, 
 Hiratsuka, Kanagawa 259-1292, Japan, 
$^{8}$STE Laboratory, Nagoya University,
   Nagoya, Aichi 464-860, Japan, 
$^{9}$National Astronomical Observatory, Tokyo 181-8588, Japan, 
$^{10}$Faculty of Science, Ibaraki University, 
   Mito, Ibaraki 310-8521, Japan, 
$^{11}$LPNHE, Ecole Polytechnique. Palaiseau CEDEX 91128, France,
$^{12}$Computational Science Laboratory, Institute of Physical and Chemical
   Research, Wako, Saitama 351-0198, Japan, and
$^{13}$Faculty of Engineering, Kanagawa University,
 Yokohama, Kanagawa 221-8686, Japan
}, 
J.~Kushida$^2$,   
S.A.~Dazeley$^3$,
P.G.~Edwards$^4$,
S.~Gunji$^5$, S.~Hara$^2$,  
T.~Hara$^6$, J.~Jinbo$^7$, 
T.~Kifune$^1$, 
H.~Kubo$^2$, 
Y.~Matsubara$^8$, 
Y.~Mizumoto$^9$, M.~Mori$^1$, 
M.~Moriya$^2$, 
H.~Muraishi$^{10}$, Y.~Muraki$^8$, 
T.~Naito$^6$, K.~Nishijima$^7$, 
J.R.~Patterson$^3$, M.D.~Roberts$^1$, 
G.P.~Rowell$^1$, T.~Sako$^{8,11}$, 
K.~Sakurazawa$^2$, Y.~Sato$^1$, R.~Susukita$^{12}$, 
T.~Tamura$^{13}$,  
T.~Tanimori$^2$, S.~Yanagita$^{10}$, 
T.~Yoshida$^{10}$, T.~Yoshikoshi$^1$, and 
A.~Yuki$^8$  
}
\end{center}

\begin{center}
{\large \bf Abstract\\}

\end{center}
\vspace{-0.5ex}
%
%

We have been successful in developing light and durable 
 mirrors made of  CFRP (Carbon Fiber Reinforced Plastic) laminates 
 for the reflector  of the new CANGAROO-II 7~m telescope.
 The reflector has a parabolic shape
 (F/1.1)  with a 30~m$^2$ effective area which 
 consists of 60 small spherical mirrors of CFRP laminates.
The orientation of each mirror can be remotely adjusted by stepping motors.
After the first adjustment work,  the reflector offers a 
 point image of about 0.$^\circ$14 (FWHM) on the optic axis.
%
\vspace{1ex}

%
%
\section{Introduction}
\label{intro.sec}

The CANGAROO team has just commenced operation of 
a new imaging Cherenkov telescope of 7~m reflector 
for observation of very high energy gamma-rays.
 With an effective collecting area of 30~m$^2$,  about three times as large as 
 the CANGAROO-I telescope of 3.8~m diameter  reflector,
 the new telescope aims at the detection of several hundreds GeV gamma-rays.
The total design of the telescope is reported elsewhere 
 (Tanimori {\it et al.} 1999), 
here we report the characteristics of its reflector.

\section{Mechanical Design}

A whole view of the reflector is shown in Figure~1. 
The reflector is an F/1.1 paraboloid with a diameter of 7.2~m.
In order to use timing information to reject night sky background, 
 we chose a parabolic shape where the dispersion of photon arrival times
 caused by different light paths in the mirror is small.

\begin{figure}[h]
\begin{center}
\label{fig:reflector}
\epsfig{file=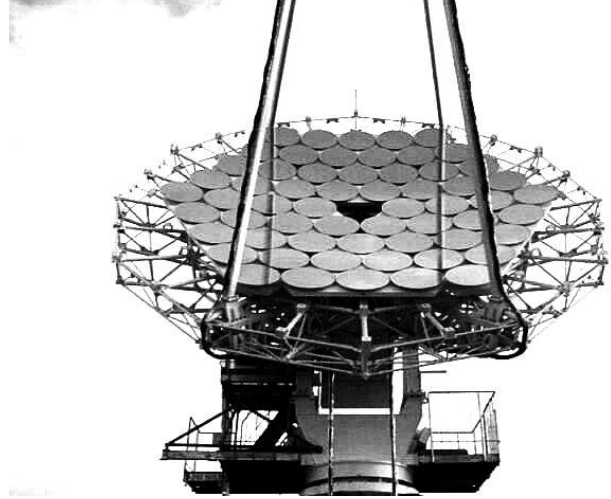, width=9.cm}
\caption{
A photo of the CANGAROO-II paraboloid reflector.
Sixty small spherical mirrors are installed on the supporting frame
with honeycomb panels.}
\end{center}
\end{figure}

Sixty spherical mirrors, each of which has an 80~cm diameter and 
a curvature radius between 16--17~m, 
 were arranged according to their curvature radii 
 from the inner to the outer sections of the reflector, 
with the shorter focal length mirrors innermost.
In the prime focal plane, there is a multi-pixel camera
with 0.$^\circ$12 pitch covering about 3 degrees of FOV
(Tanimori {\it et al.} 1999, Mori {\it et al.} 1999).

The supporting frame of the reflector mounted by 9 honeycomb panels 
is rather light in weight to save cost and assembling labor.
Several mirrors (6--9) were installed onto a honeycomb panel and 
 the alignment of the mirrors in 
 each panel was roughly adjusted with a laser beam 
 before the shipping to Australia. 
 With these adjustments, we checked our remote adjustment system as well as 
 it saved on-site labor described in Section~4.

The structure was designed to be sustained at the average velocity 
 30~km/hr of the wind load,  
 and gravitational deformations were measured to cause as small as 1$\prime$
 of deviation  at the focal plane. 
The present support frame allows us to extend the reflector up to 10~m diameter
 with additional 54 mirrors, 
 and the extension is to be completed by the beginning of 2000.

\section{Small Spherical Mirror}
\label{small_mirror.sec}

The small spherical mirrors are of 80~cm in diameter 
 and weigh only about 5.5~kg.
 Sheets of CFRP and adhesives were laid on a  metal mold,  
 sandwiching a core of low density, high shear strength foam ($rohacell$) 
 to avoid 
 twisting deformations,  and a polymer sheet coated with laminated aluminum
 was applied on the top of the layers as a reflecting material.
 The laminates were vacuum bagged and cured in an autoclave pressure vessel.

 We examined possible deterioration 
  by repeating 200 cycles of rather extreme changes in temperatures of 
 0$^\circ$--50$^\circ$C.  Change in the curvature was found to be negligible 
 after the examination.  

For protection against dust, rain, and sunshine, 
 the mirror surface was coated with fluoride.
The coat maintains more than 80~\% reflectivity down to 300~nm, 
corresponding both to the atmospheric 
 transmission cut-off at 300~nm of Cherenkov light, and to the spectrum response

 of the PMT photo-cathode with UV-transparent window (Hamamatsu, R4124UV).
 It was confirmed in a year-time-scale that 
 the reflectivity repeatedly recovers easily by washing.

The curvature radii of the mirrors was found to distribute between 
 15.9--17.1~m, with an average of 16.45~m. The mirrors
 were arranged on the support according to their radii 
 to make a smooth $f$=8~m paraboloid, and the individual facets were adjusted 
 toward the focal point by the method described later.
The image size of each mirror was measured with a light source 5.8~km away.
 A typical size is 0.$^\circ$08 (FWHM), 
 and 50~\% of the photons concentrates within $\sim$0.$^\circ$1 $\phi$.
 Including some mirrors of moderate image size, 
 the average of focusing properties of the 60 mirrors is 
 about 0.$^\circ$1 (FWHM).

\section{Remote Adjustment of the Alignments} 
\label{adjust.sec}

Two watertight stepping motors are installed at the back surface of each mirror, 
and the orientation of a mirror can be remotely adjusted in two perpendicular directions.
The minimum step size corresponds to about 1~$\times~{\rm 10}^{-4}$ degree 
at the focal plane,  
 and $\pm$~3-degree adjustment is possible.  The accuracy of 1~$\times~{\rm 10}^{-3}$ 
 degree is  retained when motors are switched off.
All mirrors are adjusted one by one using two motor drivers with relay switches
 controlled by a computer.

After the mechanical assembly of the telescope, 
we adjusted the orientations of the mirrors on-site 
 using a point light source at 5.8~km distant at night. 
 The telescope was precisely pointed at the direction of the light source 
 determined by a survey.
 All small mirrors were covered but one,  and its image on a screen 
 at the focal plane was monitored by a CCD camera installed at the 
 center of the reflector.
 The orientations of the mirrors were adjusted 
 by moving stepping motors using feedback information from 
 CCD images,  
 so that the image center 
 should lay at the focal point.

As a result of the first adjustment work, the deviation of the small mirror axis 
orientations is 0$^\circ$.03~$\pm$~0$^\circ$.01 (statistical error only) on average,  
 a larger value than expected.
 The deviation was mainly caused by temporal fluctuations of the CCD camera 
 geometry  over different nights we applied the adjustment,
 leading to deviations of the focal point on the screen.
 Removing this effect, it is estimated that the orientations can be adjusted 
 within an error of 0$^\circ$.01.

\section{Star Image, On and Off Axis}
\label{star.sec}

\begin{figwindow}[1,r,%
{\mbox{\epsfig{file=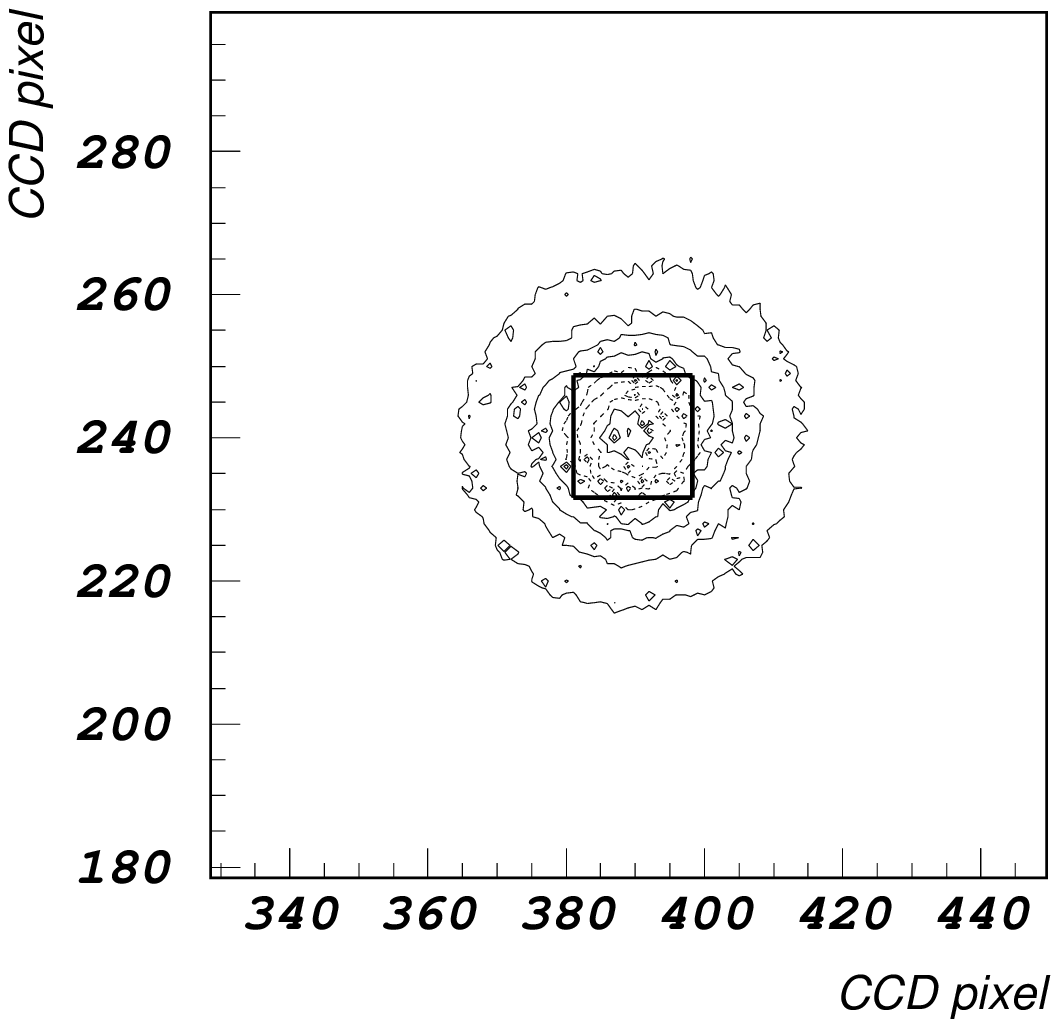,width=9.5cm}}}, 
 {A CCD image of Canopus on the optic axis. The axes are in unit of CCD pixels, 
 corresponding to a 6.7$\times$10$^{-3}$ degree pitch.
 $z$ axis is in arbitrary unit.  A square overdrawn is a scale of 
 a pitch of the camera (0$^\circ$.12 square).}]
 The focusing property of the reflector in total was measured using images of 
 several  stars tracked by the telescope.
 Images on the focal plane screen were taken by a CCD camera at the reflector 
 center.  

 In Figure~2, 
 an image of Canopus on the optic axis is shown in units of CCD pixels.
 A pixel corresponds to 6.7$\times$10$^{-3}$ degree. 
 One pixel of the camera (0$^\circ$.12 square) is overdrawn for scaling.
 A point spread function is also shown in Figure~3~(a). 
 An image size of 0$^\circ$.14$~\pm~$0$^\circ$.01 (FWHM) is deduced, and 
 30$~\pm~$4~\% of the photons concentrates in a single camera pixel.
 The image concentration is only 60~\% of the expected value in design, 
 and this dispersion might be mainly due to 
 our on-site adjustment of the mirror-orientations.

When incident rays are not parallel to the optic axis of the reflector, 
 the overall shape of the composite reflector causes aberrations.
Among them, the effect of coma aberration is dominant for a parabolic mirror
 and other aberrations of each facet mirror can be neglected.
The aberration is rather serious for a Cherenkov imaging 
 telescope since 
 a relatively wide field of view ($\sim$3$^\circ$) is needed for image analyses of 
atmospheric showers.

\end{figwindow}

\begin{figure}[ht]
\begin{center}
\label{fig:onoff}
\epsfig{file=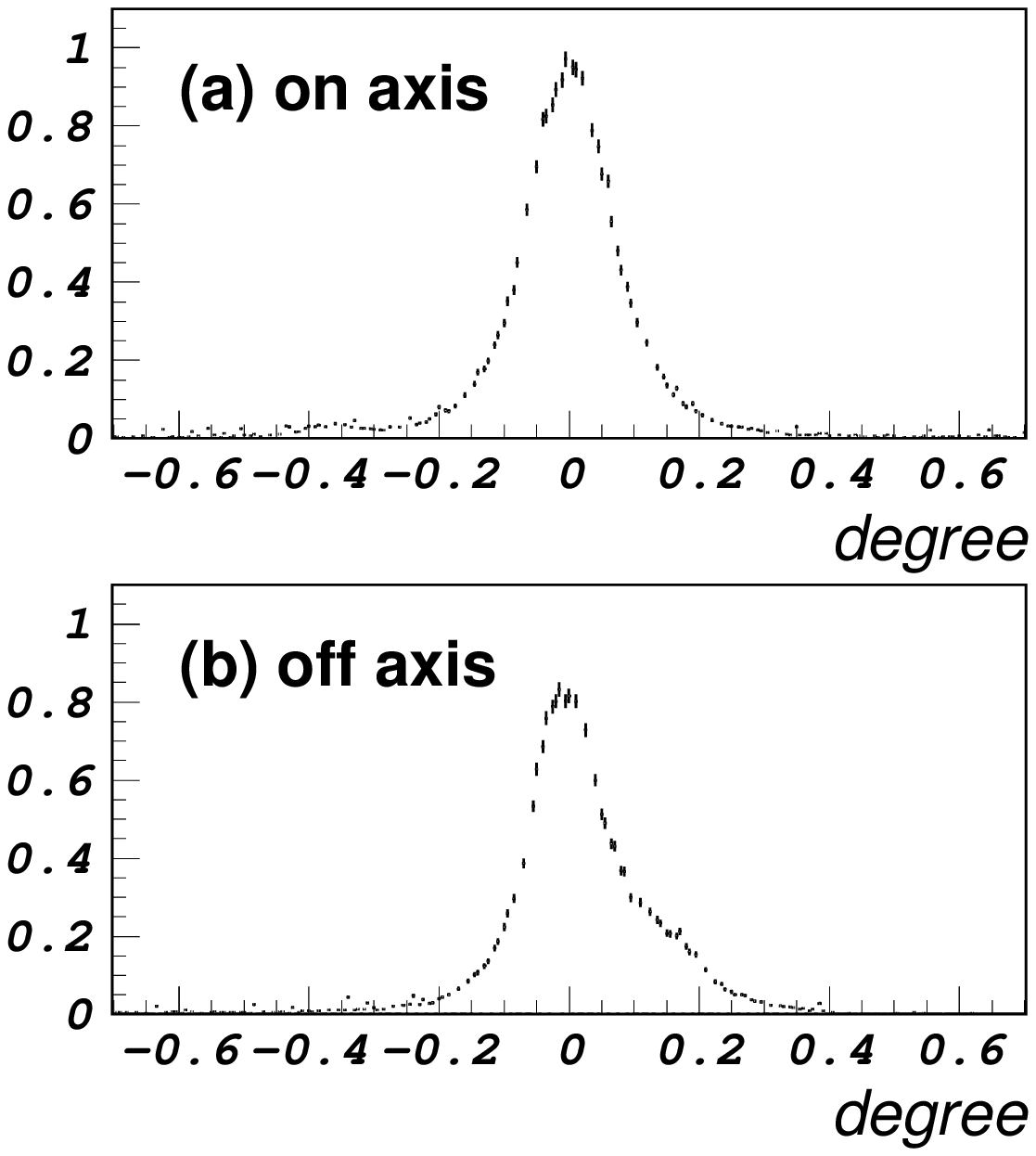,height=9.cm}
\caption{
 Measured radial point spread functions on the optic axis (a) 
 and 1$^\circ$.25 off the axis (b) are shown in a unit of degree.
 The zeros of the abscissa  are arbitrary, 
 and vertical scales are normalized by the peak height of (a).
}
\end{center}
\end{figure}

We compared off-axis images of Sirius 
 displacing the pointing coordinates by $\pm$~1.3 degree both in right ascension 
 and in declination.
 As a result of symmetric configuration and alignments,
 displacement to all the directions caused deformations consistent with each other.

Figure~3 
 shows radial point spread functions of the star 
 pointed on and 1.25 degrees off the optic axis.
 An effect of aberrations arise at $\sim$0.$^\circ$07  distant 
 from the peak center 
 (the half-width of a camera pixel is 0.$^\circ$06), 
 thus FWHMs of the on and off-axis images (about 0.$^\circ$14) are equal.
 Measurements of the concentration, however, 
 differ by about 18~\% at the edge of FOV.

The effect of gravitational deformations on the reflector 
 was measured by comparing the images of stars taken at 
 various elevation angles of the telescope.
 For elevation angles between 15--70 degrees,  the images show no dependence 
 on the elevations either in shape or in size.
 Thus the deformations are confirmed to be negligible.

\section{Summary:}

 The new CANGAROO-II 7~m telescope has been completed and 
 operations has begun. The reflector, F/1.1 paraboloid, 
 has a point spread function of 0.$^\circ$14 (FWHM) over 3 degrees of FOV, 
 with $\sim$18~\% loss of light by aberration at the FOV edge.
 About 30\% of the optical light detected from an on-axis 
 point source falls in a single camera pixel.

\section{Acknowledgments}
 The small mirrors of CFRP laminates have been developed in collaboration with 
 Mitsubishi Electric Corporation, Communication Systems Center.

%
\vspace{1ex}
\begin{center}
{\Large\bf References}
\end{center}
%
Mori, M.\ et al., 1999, Proc. 26th ICRC (Salt Lake City, 1999)\\
Tanimori, T.\ et al., 1999, Proc. 26th ICRC (Salt Lake City, 1999)
\end{document}